# White Paper on Ultra-High Energy Cosmic Rays

A. V. Olinto (University of Chicago, Chicago, IL, olinto@kicp.uchicago.edu); J. H. Adams (NASA MSFC, NSSTC, Huntsville, AL); C. D. Dermer (U.S. Naval Research Laboratory, Washington, DC); J.F. Krizmanic & J.W. Mitchell, (Goddard Space Flight Center, Greenbelt, MD); P. Sommers (Penn State University, University Park, PA); T. Stanev (Bartol Research Institute, University of Delaware, Newark, DE); F.W. Stecker (Goddard Space Flight Center, Greenbelt, MD); Y. Takahashi (University of Alabama in Huntsville, Huntsville, AL)

367 *Supporting Scientists from* http://uhecr.uchicago.edu/

J. A. Abraham, P. T. Abreu, B. S. Acharya, R. Agarwal, E.-J. Ahn, M. A.-S. Al-Seady, I. F.M. Albuquerque, V. Alekseenko, D. Allard, G. Ambrosi, M. Ambrosio, L. Anchordoqui, J. C. Anjos, C. Aramo, H. Arjomand Kermani, F. Arqueros, K. Asakimori, A. Ashtekar, H. G. Asorey, P. Assis, J. Aublin, M. Ave Pernas, S. K. Badr, N. F. Barghouty, M. G. Baring, S. L.C. Barroso, R. Battiston, G. Battistoni, B. M. Baughman, J. J. Beatty, A. D. Bellétoile, S. Y. BenZvi, A. A. Berlind, M. E. Bertaina, X. Bertou, A. Bhadra, P. L. Biermann, C. Bleve, J. Bluemer, C. Bonifazi, A. S. Borisov, K. Boruah, S. Bottai, J. T. Brack, T. J. Brandt, B. Bruna, N. G. Busca, A. M. Bykov, K.S. Caballero Mora, M. O. Calvao, L.I. Caramete, R. Caruso, M. Casolino, A. Castellina, O. Catalano, L. Cazon, J. M. Chirinos, A. S. Chou, J. Chudoba, R. W. Clay, A. Cordier, S. Coutu, C. E. Covault, H. J. Crawford, S. E. Csorna, J. C. D'Olivo, D. D'Urso, S. Dagoret-Campagne, R. Dallier, G. Dariusz, P. Datta, P. Datta, P. Davoudifar, B. R. Dawson, R. M. de Almeida, M. De Domenico, S. J. de Jong, J. R.T. de Mello Neto, I. De Mitri, V. de Souza, L. G. Dedenko, L. del Peral, O. Deligny, V. G. Denisova, S. Desai, J. V. Dias de Deus, P. N. Diep, C. Dobrigkeit, K. Dolag, L. Dorman, M. T. Dova, M. A. DuVernois, V.C. Dwivedi, T. Ebisuzaki, M. A. El-Borie, R. Engel, A. D. Erlykin, C. O. Escobar, M.C. Espirito-Santo, P. Facal San Luis, H. R.M. Falcon, A. M. Farahat, D. Fargion, G. Farrar, S.J. S.J. Fatemi, A. Fernandez-Tellez, F. Ferrer, E. Fiandrini, B. Fick, V. V. Fidelis, A. Filevich, J. P. Finley, E. D. Fokitis, E. S. Fraga, F. Fraschetti, P. Galeotti, D. Garcia-Pinto, A. Garyaka, H. E.H. Gemmeke, A. Geranios, P.L. Ghia, M. G. Giammarchi, N. Giglietto, M. Giller, H. D. Glass, H. Goldberg, P. Gonçalves, P. Gorodetzky, U. D. Goswami, E. W. Grashorn, P. K.F. Grieder, M. Grigat, C. Grupen, F. Guarino, N. Götting, D. Harari, J. L. Harton, N. Hasebe, A. Haungs, T. Hebbeker, P. Homola, D. Hooper, T. Huege, P. H. Huentemeyer, J. R. Hörandel, G. Imponente, S. Inoue, A. Insolia, F. Ionita, M. Iori, F. Kajino, K.-H. Kampert, R. P. Kane, H. Kang, P. H. Kasper, S. C. Kaushik, Y. Kawasaki, B. Kegl, B. Keilhauer, E. Kemp, J. Kempa, M. Khakian Ghomi, R. M. Kieckhafer, D. B. Kieda, A. A. Kirillov, H. O. Klages, M. Kleifges, S. R. Klein, R. Knapik, J. Knapp, D.-H. Koang, K. Kotera, A. Kryemadhi, F. G.I. Kuehn, A. P. Kumar, A. Kusenko, C. Lachaud, S. J. Lafebre, B. L. Lago, P. Le Coultre, D. Lebrun, P. L.G. Lebrun, A. Letessier-Selvon, G. Li, H. Lyberis, M.C. Maccarone, C. Macolino, K. Mannheim, P. Mantsch, I. C. Maris, O. M. Martinez, R. Martirosov, K. Mase, Y. Matsubara, A. Mattei, J. M. Matthews, G. Matthiae, S. C. Mavrodiev, P. O. Mazur, T. McCauley, T.J.L. McComb, M. A. McEwen, G. A. Medina-Tanco, A. Meli, E. A. Menichetti, P. Meszaros, A. A. Mikhailov, T. A. Mir, L. Miramonti, R. K. Mishra, S. K. Mishra, A. K. Mitra, I. Mocioiu, M. Monasor, D. Monnier Ragaigne, F. Montanet, I. V. Moskalenko, M. A. Mostafa, C. A. Moura, M. Nagano, D. V. Naumov, V. A. Naumov, L. Nellen, N. M. Nesterova, D. F. Nitz, D. Nosek, A. Okumura, A. C. Olinto, S. Osone, N. Pacheco, G. Parente, E. Parizot, B. Pattison, T. Paul, V. Pavlidou, M. Pearce, R. Pelayo, I. M. Pepe, L. Perrone, S. Petrera, A. Petrolini, Y. V. Petrov, A. A. Petrukhin, P. Picozza, T. Pierog, M. J.M. Pimenta, O. Pisanti, E. Plagnol, N. J. Poplawski, P. Privitera, M. Prouza, V. Ptuskin, J. J. Quenby, J. Rautenberg, D. Ravignani, P. Raychaudhuri, S. Razzaque, P. J. Reardon, H. Rebel, B. Revenu, S. Riggi, M. Risse, G. Rodriguez, M.D. Rodriguez Frias, M. A. Roth, B. Rouillé d'Orfeuil, A. C. Rovero, S. A.-A. Saleh, J. M. Santander, A. Santangelo, E. M. Santos, F. Sarazin, F. Schmidt, T. C. Schmidt, O. Scholten, S. Schulte, D. A. Schuster, F. Schüssler, M. Scuderi, D. Semikoz, M. Settimo, A. Shalchi, R. Shanidze, R. C. Shellard, M. M. Sherif, G. Siemieniec-Ozieblo, G. H.W. Sigl, M. G. Signore, A. F. Sill, S. I. Sinegovsky, K. R. Sinha, R. Smida, G. R. Snow, P. Sokolsky, P. Spillantini, H. Spinka, A. Stamerra, S. Stoica, R. E. Streitmatter, T. Suomijarvi, M. Svanidze, J. D. Swain, B. Szabelska, J. Szabelski, Z. Szadkowski, H. Takami, Y. Takizawa, T. P.H. Tam, S. V. Ter-Antonyan, M. Teshima, B. Thompson, D. Timashkov, C. Timmermans, A. K. Tiwari, C. M. Tiwari, L. G. Tkachev, W. Toki, B. A. Tomé, A. S. Tonachini, H. S. Tornabene, P. Travnicek, R. K. Tripathi, M. J. Tueros, N. Turini, R. M. Ulrich, M. Unger, M. Urban, G. K. Ustinova, A. G. Vacchi, R. O. Vainio, A. M. van den Berg, D. Veberic, P. Velinov, T. M. Venters, L. Villasenor, T. Vo Van, L. R. von Lindern, S. Vorobiov, A. A. Watson, J. W. Watts, J. P. Wefel, T. J. Weiler, S. Westerhoff, L. R. Wiencke, H. Wilczynski, G. Wilk, R.J. Wilkes, T. L. Wilson, T. Yamamoto, I. I. Yashin, P. Younk, A. Yushkov, J. Zabierowski, E. Zas, D. Zavrtanik, A. Zepeda, B. Zhang, A. Zuccaro Marchi

A fundamental question that *can be answered in the next decade* is:
WHAT IS THE ORIGIN OF THE HIGHEST ENERGY COSMIC PARTICLES?

The discovery of the sources of the highest energy cosmic rays will reveal the workings of the most energetic astrophysical environments in the recent universe. Candidate sources range from the birth of compact objects to explosions related to gamma-ray bursts or generated around supermassive black holes in active galactic nuclei. In addition to beginning a new era of high-energy astrophysics, the study of ultra-high energy cosmic rays will constrain the structure of the Galactic and extragalactic magnetic fields. The propagation of these particles from source to Earth also probes the cosmic background radiation and gives insight into particle interactions at orders of magnitude higher energy than can be achieved in terrestrial laboratories. Next generation observatories designed to study the highest energy cosmic rays will have unprecedented sensitivity to ultra-high energy photons and neutrinos, which will further illuminate the workings of the universe at the most extreme energies. For this challenge to be met during the 2010-2020 decade, a significant increase in the integrated exposure to cosmic rays above $6 \cdot 10^{19}$ eV will be necessary. The technical capabilities for answering this open question are at hand and the time is ripe for exploring **Charged Particle Astronomy**.

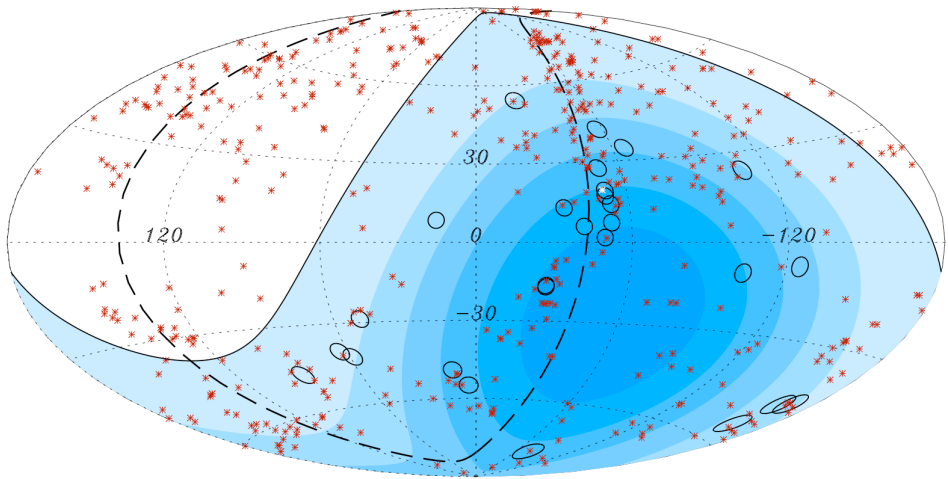

*Fig. 1: Map in galactic coordinates of AGNs in the VCV catalog with z < 0.018, D < 75 Mpc (red stars) and 27 events with $E > 5.7 \cdot 10^{19}$ eV (black $3.2^o$ circles) observed by the Auger Observatory [1]. The solid line outlines the field of view. Blue shading indicates regions of equal exposure. The dashed line is the Supergalactic plane and Centaurus A is marked in white.*

**Introduction**

Cosmic rays with energies that can exceed $10^{20}$ eV confront us with some of the most interesting and challenging questions in astrophysics: Where do they come from? How can they be accelerated to such high energies? What kind of particles are they? What do they tell us about these extreme cosmic accelerators and their large-scale structure distribution? How strong are the magnetic fields that they traverse on their way to Earth? How do they interact with the cosmic background radiation? What can we learn about particle interactions at these otherwise inaccessible energies?

Although the first detection of ultra-high energy cosmic rays (UHECRs) dates back to 1962[2], it was only during the 1990s that an international effort began to address these questions with the necessary large-scale observatories. The largest detectors operating during the 1990s were the Akeno Giant Air Shower Array (AGASA)[3], a 100 km$^2$ ground array of scintillators in Japan, and the High Resolution Fly's Eye (HiRes)[4], a pair of stereo fluorescence telescopes operated in Utah until 2006. During their lifetimes, AGASA reached an exposure of 1.6 10$^3$ km$^2$ sr yr while HiRes reached twice that.

With an exposure of 10$^3$ km$^2$ sr yr, the important question of whether the spectrum of UHECR exhibits the well-known GZK effect had contradictory answers: AGASA reported an excess flux unlike the expected suppression while early results from HiRes were consistent with the GZK prediction. The GZK effect is named after Greisen[5], Zatsepin, and Kuzmin[6] who predicted in 1966 a dramatic steepening of the spectrum above a few times 10$^{19}$ eV caused by the interaction of ultra-high-energy cosmic rays with the cosmic microwave background (CMB) as they propagate from an extragalactic source to Earth. The result of these interactions is pion photoproduction in the case of proton primaries and photo-dissociation for heavier nuclei.

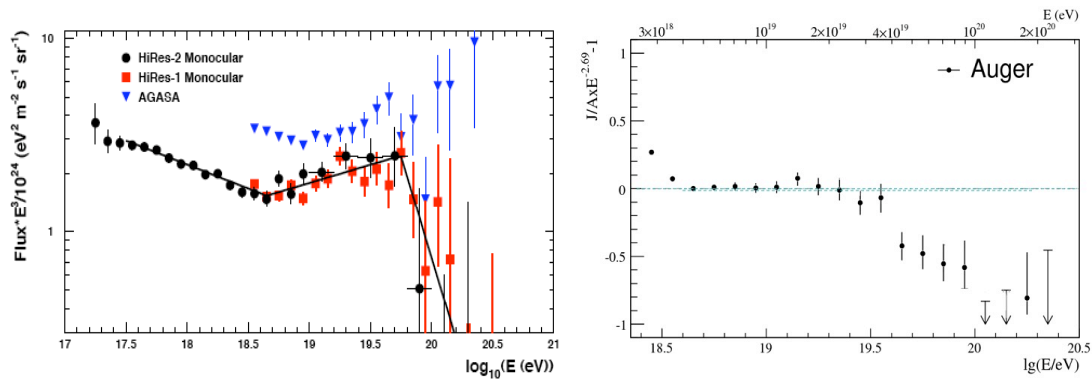

*Fig. 2. flux x $E^5$ HiRes[7] and AGASA(left); deviation from $E^{-2.69}$ Auger[8] (right).*

In 2008, HiRes[7] published the first significant observation of the GZK suppression, as displayed in Fig. 2 (left). This was confirmed by the Pierre Auger Observatory[8] based on 7 10$^3$ km$^2$ sr yr exposure, shown in Fig. 2 (right). Inaugurated in 2008, Auger is the largest observatory at present. Constructed in the province of Mendoza, Argentina, by a collaboration of 18 countries, it consists of a 3,000 km$^2$ array of water Cherenkov stations with 1.5 km spacing overlooked by four fluorescence telescopes. The combination of two techniques into a hybrid observatory maximizes the precision in the reconstruction of air showers, allowing for good control of systematics. The largest observatory in the northern hemisphere, the Telescope Array (TA)[9] is also hybrid. Situated in Utah, it covers 760 km$^2$ with scintillators spaced every 1.2 km overlooked by fluorescence telescopes.

The confirmation of the GZK feature settles the question of whether acceleration in astrophysical sources can explain the high-energy spectrum, ending the need for exotic alternatives that avoid the GZK feature. This landmark measurement also opens the way to astronomical searches for sources in the nearby extragalactic universe using the arrival directions of trans-GZK cosmic rays. Above the GZK threshold energy, observable sources must lie within about 100 Mpc, the so-called GZK horizon or GZK sphere. Fig. 3 shows the fraction of cosmic rays that arrive on Earth from a given distance for different energy protons (from 4 to 9 10$^{19}$ eV, left) and for different nuclei (protons, He, CNO, and Fe) arriving with 6 10$^{19}$ eV (right). One can define the GZK horizon as the distance from which 50% of primaries originate, which ranges from 150 Mpc to 10 Mpc in Fig.3 At these energies light composite nuclei are promptly dissociated by CMB photons, while protons and iron may reach us from sources at distances up to 100 Mpc. Matter is known to be distributed inhomogeneously within the GZK sphere, so the cosmic ray arrival directions should exhibit an anisotropic distribution above the GZK energy threshold, provided intervening magnetic fields are not too strong. At the highest energies, the isotropic diffuse flux from sources beyond the GZK radius is strongly suppressed.

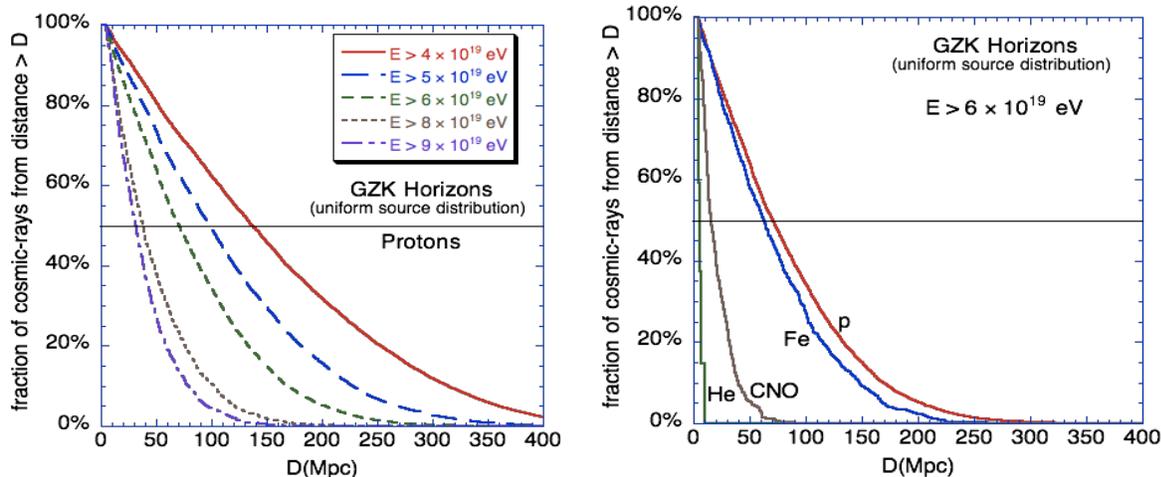

*Fig. 3: GZK Horizons of protons with 4 to 9 $10^{19}$ eV; & p, He, CNO, Fe at 6 $10^{19}$ eV[10].*

Observation of the expected anisotropy above the GZK threshold was first reported in 2007 when Auger showed that the arrival directions of 27 cosmic rays with energies above 5.7 $10^{19}$ eV exhibit a statistically significant correlation with the anisotropically distributed galaxies in the 12th Veron-Cetty & Veron [11] (VCV) catalog of active galactic nuclei (AGN) [1]. Analysis of an independent data set confirmed anisotropy at a confidence level of over 99%. No corresponding correlation was observed in the northern hemisphere by HiRes [12], which examined their 13 highest energy events and found the distribution to be consistent with isotropy at 97% probability. With the small number of events, it is not yet clear what these results may indicate about the sources or their distribution. However, the Auger observation of anisotropy suggests that above an energy of about 6 $10^{19}$ eV, *Charged Particle Astronomy* is feasible.

**Candidate Sources**

Cosmic rays can be accelerated in astrophysical plasmas when the energy from large-scale macroscopic motions, such as shocks, winds, and turbulent flows, is transferred to individual particles. Basic requirements on cosmic accelerators for reaching energies of $10^{20}$ eV are summarized in Fig. 4 in what is known as a Hillas plot[13], where the typical magnetic field B of an astrophysical object is plotted versus its size L. The maximum energy of accelerated particles, $E_{max}$, can be estimated by requiring that the gyroradius of the particle be contained in the acceleration region: $E_{max} < Z\,BL$, where Z is the charge of the particle. These extreme requirements leave a small number of known types of objects as possible candidates that range in size from neutron stars to cluster shocks. The Auger AGN correlation argues that some sources cannot be much further than about 100 Mpc, which rules out rare and distant sources, such as massive clusters of galaxies and the most powerful radio galaxies. With only 27 events it is difficult to reach any definite conclusion, but there seems to be a number of events in the neighborhood of Centaurus A, while M87 is tantalizingly quiet (see Fig. 1).

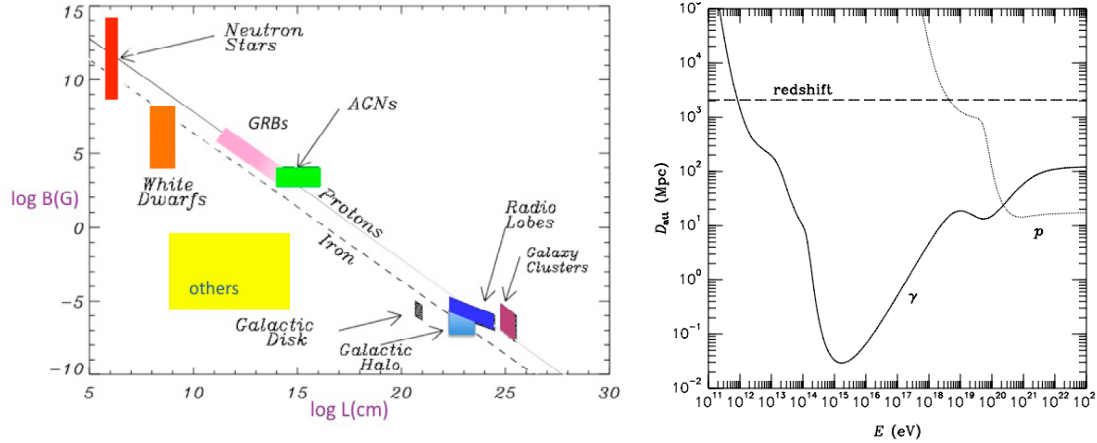

*Fig. 4: Hillas Plot for $10^{20}$ eV p & Fe.   Fig 5. p & γ attenuation off CMB, IR, radio[14].*

Although the correlation with the VCV catalog has stimulated much activity in models based on AGN acceleration, the observed correlation does not prove that AGNs are the UHECR sources. Since the position of active galaxies in the VCV catalog are themselves correlated with the large-scale distribution of matter within the GZK sphere, any type of source that displays this overall distribution is a plausible candidate. The Auger trans-GZK events also correlated with PSCz sources [15], with HI emitting galaxies[16], and Swift hard X-ray sources[17]. These catalogs also trace the large-scale matter distribution in the local universe. The sources could even reside in ordinary galaxies, especially if they are transient, such as gamma-ray bursts.

A combination of large sky exposure and precise spectrum and composition measurements is required to discover the cosmic accelerators responsible for these extreme particle energies. The diffuse spectrum convolves source injection with source density evolution and propagation through cosmic backgrounds. More sensitive overall spectral measurements will help determine the source density and evolution, and may test fundamental physics. Theories of the nature of space-time approaching the Planck scale can admit or imply violations of Lorentz invariance, which may modify the GZK effect above $10^{20}$ eV [18].

Competing models of the cosmic accelerators will be best tested when we precisely measure the cosmic ray spectrum of individual sources. This is now known to be feasible for a cosmic ray observatory, limited only by the ***amount of exposure*** to each source.

**Multimessenger Studies**

Understanding the production and propagation of cosmic rays has ramifications for all aspects of high-energy astrophysics. The sources of UHECRs are likely to emit energetic neutrinos and gamma rays that may be observed by present and future observatories. With the GZK effect firmly established in the contemporary universe and with knowledge of the cosmic ray spectrum and composition, cosmological models give clear expectations for the spectrum of diffuse neutrinos which should have accumulated over cosmic time. Neutrino observations are especially important in the $10^{15}$ eV energy range since the cosmic microwave background excludes extragalactic gamma rays above $10^{14}$ eV (Fig. 5). The IceCube Observatory and other large neutrino detectors of the future will have the capability to measure the lower energy part of the GZK neutrino spectrum, which results from neutron beta decays, while radio and acoustic techniques (e.g., RICE, ANITA) are able to probe the high energy end. Neutrino observatories may also measure astrophysical neutrinos coming straight from the hearts of cosmic ray sources themselves. UHECR observatories can detect a few GZK neutrinos per year above ~$10^{18}$ eV. The Auger Observatory has recently announced the most stringent tau neutrino flux limit for energies ~ $10^{18}$ eV to $10^{20}$ eV[19] (Fig.6), while ANITA announced the strongest limit on three flavors ~ $10^{19}$ eV to $10^{22}$ eV[20] (Fig.7).

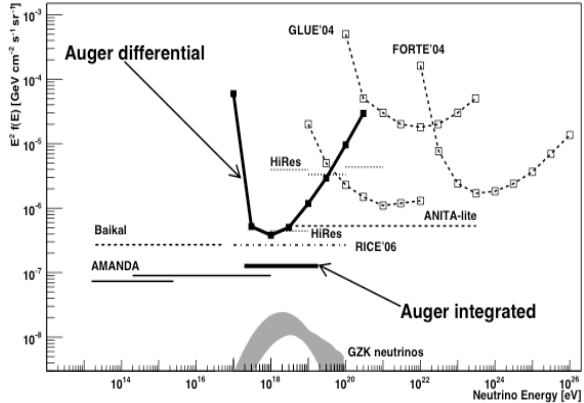 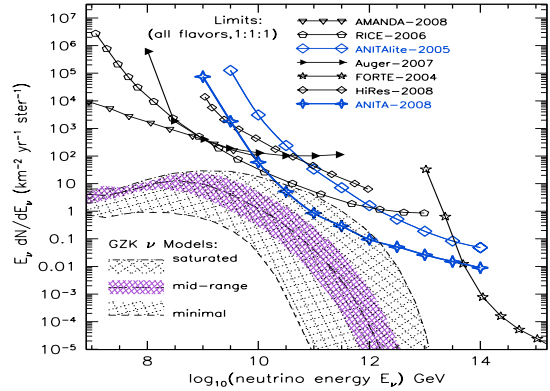

*Fig. 6: Auger tau neutrino limit[19].     Fig. 7: ANITA limits on 3 flavor neutrinos[20].*

Gamma-ray astronomy has two windows of opportunity to help determine the origin of UHECRs. Below $10^{14}$ eV, where photon attenuation by cosmic background radiation is not significant (Fig. 5), high-energy gamma ray observatories will provide invaluable detailed information about candidate sources. The Fermi Gamma-ray Space Telescope offers a new look at sources of gamma rays from ~$2 \cdot 10^{7}$ to $3 \cdot 10^{11}$ eV. Above these energies, VERITAS, MAGIC, and HESS are revealing new details about sources of gamma rays from ~$3 \cdot 10^{10}$ eV to $3 \cdot 10^{13}$ eV. Plans for HAWC, AGIS, and CTA offer exciting prospects for future multi-messenger studies.

A new gamma-ray window that will be opened with large aperture UHECR observatories is the direct detection of photons at energies above the CMB attenuation, above ~$10^{19}$ eV. Auger South has demonstrated great potential to search for photons among cosmic rays, by setting the best limits on the fraction of photons at these energies[21]. The lack of observation of photons or neutrinos strongly disfavors top-down models. The decay of GZK neutral pions gives an expected photon flux, which should be detectable with proposed larger aperture UHECR observatories.

The cosmic ray spectrum and composition at lower energies will also help us to understand the origin of cosmic rays by studying the transition between Galactic to extragalactic cosmic rays. Projects such as the Telescope Array Low Energy Extension (TALE), and the Auger South enhancements, AMIGA (Auger Muons and Infill for the Ground Array) and HEAT (High Elevation Auger Telescopes), show great promise in making these accurate measurements.

**Magnetic Fields**

UHECRs are charged particles and thus provide a way to measure integrated magnetic fields: the deflection angle relative to a known source position is inversely proportional to energy. Protons with energies above $6 \cdot 10^{19}$ eV traversing magnetic fields of $10^{-6}$ G over a kpc (or $10^{-9}$ G over a Mpc) experience deflections of order $1°$. To study our magnetic environment in all directions in this way, it is necessary to measure cosmic ray deflections in all parts of the sky. The combination of full-sky measurements and large aperture can map out nearby sources and determine the typical deflections from source to Earth. The correlation of arrival directions with AGNs out to 75 Mpc already implies new limits on average intergalactic magnetic field strengths. With increased exposure, trans-GZK cosmic rays offer an important probe of the magnetic environment in our region of the cosmos, with implications for cosmology and astrophysics.

**Opportunities for the Next Decade**

The 2010-2020 decade should be remembered as the time when charged particle astronomy was developed. This new window is expected to open at ultra-high energies due to the weakening effect of cosmic magnetic fields combined with the limited volume containing observable sources. This expectation was recently verified with the detection of particles with energies above $6 \times 10^{19}$ eV that appear to preferentially follow the distribution of AGN within 75 Mpc from Earth. To fully explore this new view of the universe, next-generation observatories need to be built that observe the full sky and can reach an order of magnitude increase in exposure. Auger South will add about $7 \times 10^3$ km$^2$ sr yr each year in exposure in the south with a total accumulation reaching about $10^5$ km$^2$ sr yr at the end of the decade. TA will add about $2 \times 10^4$ km$^2$ sr yr to the northern exposure. Given the number of trans-GZK events observed by Auger South thus far, an estimate of astronomically useful events that current observatories could detect by 2020 would be less than 500. For comparison, the number of AGNs in the VCV catalog within 100 Mpc is about 700. Clearly progress to unveiling the sources of UHECRs will be hampered by the lack of enough exposure to even the nearby sources.

Large aperture projects are being proposed to meet the need for full sky coverage and increase in exposure over the next decade: the ground-based Auger North Observatory[22] and the space-based JEM-EUSO[23], OWL[24], and Super-EUSO[25]. The Pierre Auger Collaboration has proposed to build the northern observatory in the southeast corner of Colorado that would increase the number of observed trans-GZK events to about 2,000 over the next decade. Auger North will use the same techniques as Auger South and will provide a complementary view of the sky optimized to observe at trans-GZK energies. The spacing between tanks of the 20,000 km$^2$ array will be 2.3 km, providing full aperture at $6 \times 10^{19}$ eV and above. The design calls for full fluorescence coverage to ensure accurate energy, direction, and composition measurements for trans-GZK events.

To reach even larger instantaneous apertures, the alternative to ground observatories is to deploy dedicated observatories in space that can observe UHECR showers in the atmosphere by looking down toward the Earth. The Extreme Universe Space Observatory on the Japanese Experiment Module (JEM-EUSO) is being planned for deployment on the International Space Station. JEM-EUSO uses a near-UV telescope with 2.5 m diameter and 60$^\circ$ field-of-view to detect fluorescence from UHECR events. JEM-EUSO may detect ~1,000 particles above $7 \times 10^{19}$ eV in a three year mission.

The Orbiting Wide-Angle Light Collectors (OWL) mission employs two telescopes separated by ~ 600 km in 1,000 km near equatorial orbits to stereoscopically image fluorescence from UHECR. Developed in formal NASA instrument and mission studies, the baseline 3 m optical-aperture UV telescopes launch on a single conventional vehicle. OWL reaches full aperture at $10^{19}$ eV and a 5 yr OWL mission would deliver ~$10^6$ km$^2$ sr yr. Super-EUSO is a European-led effort also based on a free-flyer mission.

The time is ripe to explore the hints given by Auger South that there are sources of UHECRs that can be identified with a significant increase in exposure to trans-GZK events. Auger North is now in the R&D phase and can start construction by 2010. JEM-EUSO could fly as soon as 2013. This decade may also witness the launch of OWL and Super-EUSO and the development of radio and acoustic detection techniques. With a coordinated effort, the next generation observatories can be built to explore some of the ~ 5 million trans-GZK events the Earth's atmosphere receives per year.

*"Whatever happens, in a subject where the dullest and most conventional theories involve massive, spinning, black holes, ultra-relativistic blast waves and 100 GT fields threading nuclear matter, the future is guaranteed to be interesting."*
– *Roger Blandford 1999 [26]*

**References**
[1] Pierre Auger Collaboration (2007) *Science* 318 (5852), 938
[2] J. Linsley (1963) *Phys. Rev. Lett*. 10 146
[3] http://www-akeno.icrr.u-tokyo.ac.jp/AGASA/
[4] http://www.cosmic-ray.org/
[5] K . Greisen (1966) *Phys. Rev. Lett*. 16 748
[6] G .T. Zatsepin and V. A. Kuzmin (1966) *Zh. Eksp. Teor. Fiz*. 4 114, *JETP Lett.* 4 78
[7] HiRes Collaboration (2008) *Phys.Rev.Lett*.100:101101
[8] The Pierre Auger Collaboration (2008) *Phys. Rev. Lett*. 101, 061101
[9] http://www.telescopearray.org/
[10] D. Allard et al. (2008) *JCAP*10, 033.
[11] M.-P. Véron-Cetty and P. Véron, (2006) *Astron. Astrophys*. 455 773
[12] HiRes Collaboration (2008) Astropart.Phys.30:175-179
[13] A. M. Hillas (1984) *Ann. Rev. Astron. Astrophys*. 22, 425
[14] S. Lee (1998) *Phys.Rev.D*58 043004
[15] T. Kashti and E. Waxman (2008) *JCAP* 0805:006
[16] G. Ghiselini et al. (2008) arXiv:0806.2393
[17] M. R. George et al. (2008) *MNRAS* 388L, 59G
[18] S. T. Scully and F. W. Stecker (2009) *Astropart. Phys* arXiv:0811.2230.
[19] Pierre Auger Collaboration (2008) *Phys. Rev. Letters* 100, 211101
[20] ANITA collaboration (2009) arXiv:0812.2715
[21] Pierre Auger Collaboration (2008) *Astropart. Phys.* 29, 243-256
[22] www.augernorth.org
[23] Y. Takahashi et al. (2007) *J. Phys. C.* 65, 012022; http://jemeuso.riken.jp/.
[24] F. W. Stecker et al. (2004) *Nucl. Phys. B*, 136C, 433; http://owl.gsfc.nasa.gov/
[25] A. Santangelo et al. (2005) Proceedings of the 39th ESLAB Symposium
[26] R. Blandford (1999) *Lect. Notes Phys*. 530, 1